\documentclass[aps,pre,twocolumn,showpacs]{revtex4}

\usepackage{graphicx}

\newcommand{\be}{\begin{equation}}
\newcommand{\ee}{\end{equation}}
\newcommand{\bea}{\begin{eqnarray}}
\newcommand{\eea}{\end{eqnarray}}

\newcommand{\br}{\mathbf{r}}

\newcommand{\bt}{\mathbf{t}}

\newcommand{\bxi}{\mbox{\boldmath${\xi}$}}

\newcommand{\md}{\mathrm{d}}
\def\eq#1{Eq.~(\ref{#1})}

\begin{document}

\title{Strand diffusion-limited closure of denaturation bubbles in DNA}

\author{Anil Kumar Dasanna}
\author{Nicolas Destainville}
\author{John Palmeri}
\author{Manoel Manghi}
\email{manghi@irsamc.ups-tlse.fr}
%\shortauthor{A. K. Dasanna \textit{et al.}}

\affiliation{                    
  (1) Universit\'e de Toulouse; UPS; Laboratoire de Physique Th\'eorique (IRSAMC); F-31062 Toulouse, France, EU\\
  (2) CNRS; LPT (IRSAMC); F-31062 Toulouse, France, EU
}
\pacs{87.15.H-,87.15.A-,82.39.Pj}

\begin{abstract}
The closure dynamics of a pre-equilibrated DNA denaturation bubble is studied using both Brownian dynamics simulations and an analytical approach. The numerical model consists of two semi-flexible interacting single strands (ssDNA) and a bending modulus which depends on the base-pair state, with double-strand DNA (dsDNA) segments being 50 times stiffer than ssDNA ones. For DNA lengths from $N=20$ to 100 base-pairs (bp) and initial bubble sizes of $N-6$~bp, long closure times of 0.1 to 4 $\mu$s are found, following a scaling law in $N^{2.4}$. The bubble starts to close by a fast zipping which stops when the bubble reaches a highly bent metastable state of length around 10 bp. The limiting final step to complete closure is controlled by the dsDNA ``arms'' rotational diffusion, with closure occurring once they are nearly aligned. The central role of chain bending, which cannot be accounted for in one-dimensional models, is thus illuminated.
\end{abstract}

\maketitle

\section{Introduction}
\label{intro}

%definition of bubble closure dynamics
Understanding the dynamics of biological processes such as transcription, duplication or DNA translocation by viruses is a challenge for biophysicists. By itself, the dynamics of double-stranded DNA (dsDNA) is a complex issue which has been tackled at two different scales~: (i)~at the macromolecule scale, dsDNA is a model semi-flexible polymer whose internal structure is ignored and whose dynamics is controlled by thermal fluctuations and bending modes~\cite{maggs,goldstein}; (ii)~at the base-pair scale, the dynamics focuses either on the base-pair closure during DNA renaturation at room temperature, the so-called DNA zipping~\cite{sikorav}, or on the breathing dynamics, i.e. the fast opening and subsequent closure of small bubbles, a local opening of successive base-pairs, with very low probability~\cite{metzlerJPA,metzlerPRL,sung,collins}.
Such studies do not consider the closure dynamics of a thermalized or pre-equilibrated denaturation bubble which couples both scales, i.e. local base-pair closure and chain diffusion modes. The closure of such a denaturation bubble occurs for instance at the final stage of DNA transcription, when RNA-polymerase leaves the locally open DNA~\cite{cell}, or is observed \textit{in vitro} in DNA solutions as a rare event with the largest time scale. Altan-Bonnet \textit{et al.} measured the closure dynamics of denaturation bubbles of 18 base-pairs (bp) by fluorescence correlation spectroscopy and found surprisingly long closure time in the $20-100~\mu$s range~\cite{altan}. Such long times were also measured in bulk experiments on hairpin formation in ssDNA and RNA oligomers, which are much slower than theoretically estimated times of end-to-end contacts for simple semi-flexible polymers~\cite{ansari}. Bubble lifetimes of about $1~\mu$s have also been observed for DNA oligomers of 14~bp in NMR measurements of the imino proton exchange~\cite{warmlander}.

%other models
Several models have been used for studying bubble breathing and in attempts to explain these large experimental bubble lifetimes. (i)~The Poland-Scheraga model~\cite{PS} is a one-dimensional (1D) Ising model modified to account for the entropic penalty of creating a closed flexible loop. This term leads to a non-monotonic free energy landscape in which the typical breathing time comes from a Kramer's process~\cite{metzlerJPA,metzlerPRL,bar}. (ii)~The Peyrard-Bishop model is a non-linear phonon 1D model where bubbles emerge as soliton-like solutions of undamped Newton's equations in a Morse potential~\cite{PB,cuesta,collins}. Although these models capture the short time scale breathing dynamics, they are not adapted to the issues of renaturation and ``equilibrated'' large bubble closure since they miss DNA diffusion in solution. But strand dynamics is expected to be dominant at least for long DNAs and large bubbles, since for a flexible chain of length $N$ the Rouse diffusion time scales like~$N^2$. (iii)~Other models fit well the experimental auto-correlation function~\cite{altan,srivastava,bicout}, but with the relaxation time as a fitting parameter, which does not shed light on the origin of such large times.

%our approach
In an attempt to explain the mechanism behind these large bubble lifetimes, we focus on the out-of-equilibrium closure of a thermalized denaturation bubble using both Brownian Dynamics (BD) simulations and analytical approaches. We implement two numerical models where the different bending rigidities of dsDNA segments with a persistence length of roughly $\ell_{\rm ds}= 150$~bp and ssDNA ones with $\ell_{\rm ss}=3$~bp are explicitly included, and whose coupling with base pairing has been central to understanding equilibrium properties~\cite{PRL,JPCM}.
%
%our results
\begin{figure}[t]
\includegraphics[width=8.5cm]{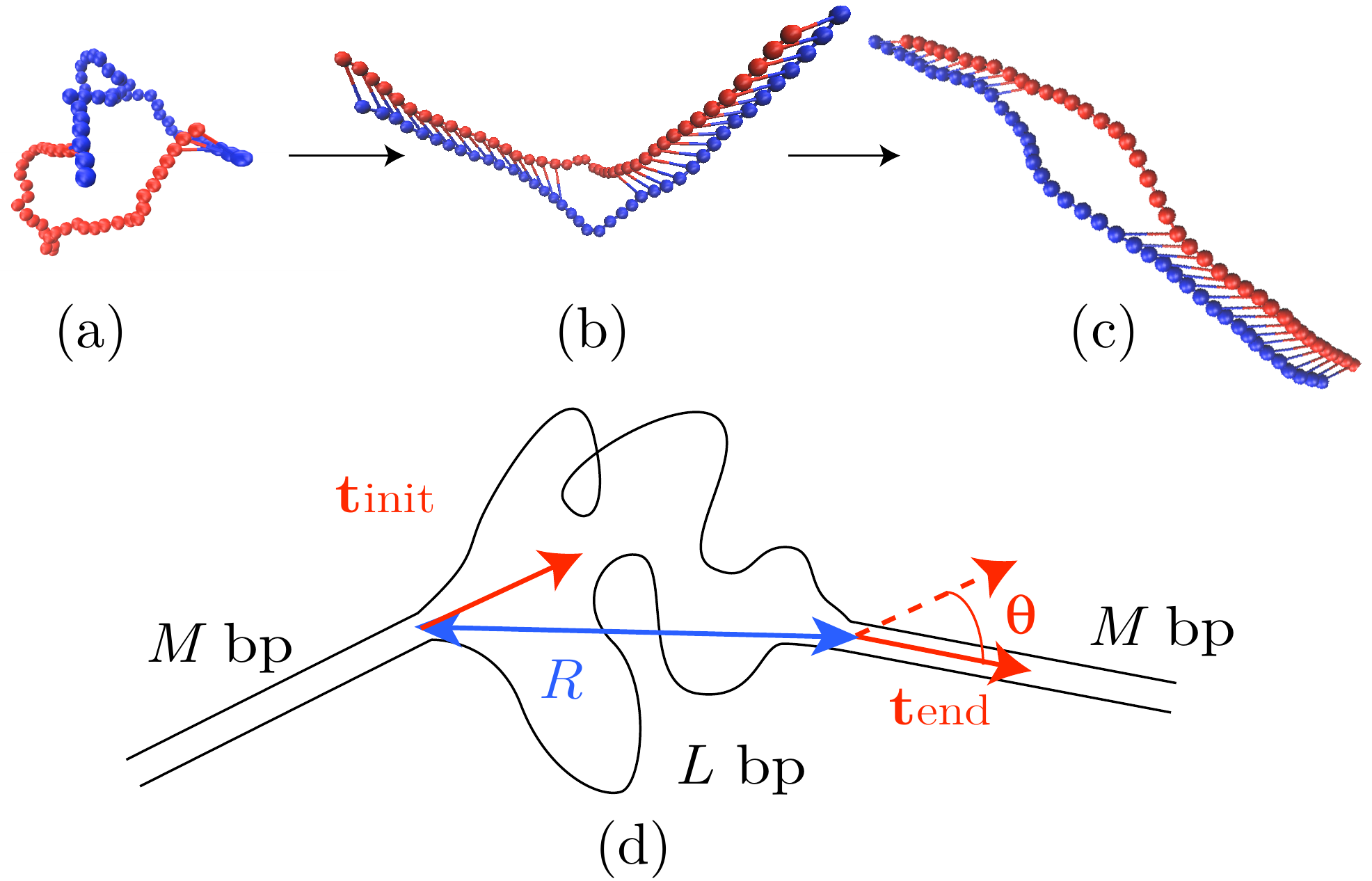}
\caption{Snapshots of a typical Brownian dynamics simulation showing (a)~the initial equilibrated bubble, (b)~the metastable state, and (c)~the bubble just before closure ($N=40$~bp). (d)~Sketch of the metastable state.}
\label{snapshots}
\end{figure}
We show that the denaturation bubble closure occurs in two steps (Figure~\ref{snapshots}). The first step consists in a fast \textit{zipping} of the initial bubble until a metastable bubble state of length $\sim10$~bp is reached. The driving force for this fast kinetics is the energetic gain in base-pair closure at room temperature. At some point, the metastable bubble is so bent that zipping becomes forbidden by the large bending energy cost to close the bubble. The second step of the closure is then controlled by the relaxation of the bent state through the diffusion of the two stiff dsDNA arms. Once the two rigid strands are aligned, bubble closure occurs almost instantaneously. The first mean passage closure time is found to scale with the DNA length $N$ (in bp units) as $\tau_{\rm closure}\sim N^{2.4\pm0.1}$ for $20\leq N\leq 100$ and mesoscopic parameters typical of real DNA.

\section{Models}

We simulated the closure of a large bubble of initial length $L_0=N-6$ in the middle of an homopolymer DNA. We used both BD simulations of two interacting semi-flexible strands, and the Kinetic Monte Carlo (KMC) algorithm which simulates the mean semi-flexible chain with an internal Ising spin dynamics corresponding to the bp state (broken or unbroken)~\cite{PRL,JPCM}.

\subsection{Brownian dynamics simulations} The DNA is modeled by two interacting bead-spring chains each made of $N$ beads located at $\br_i$. The Hamiltonian, $\mathcal{H} = \mathcal{H}_{\rm el}^{(1)} + \mathcal{H}_{\rm el}^{(2)} + \mathcal{H}_{\rm int}$, has three terms. The elastic energy of strands $i=1,2$ is
\be
\mathcal{H}_{\rm el}^{(i)} = \sum_{j=1}^{N-1}\left[\frac{\kappa_s}{2}\,(|\bt_j|-a)^2 + \kappa_{b,j}\:(1-\hat{\bt}_{j}\cdot\hat{\bt}_{j+1})\right]
\label{el}
\ee
where $\bt_j=\br_{j+1}-\br_j$ and $\hat{\bt}_j=\bt_j/|\bt_j|$. The first term of the rhs. of \eq{el} is the stretching energy with stretching modulus $\beta\kappa_{s}=100$ ($\beta^{-1}=k_{\rm B}T$ where $T$ is room temperature) and $a=0.34$~nm is the equilibrium distance between two beads in each strand. The second term is the usual bending energy with a bending modulus $\kappa_{b,j}$ that depends on the local chain configuration ($\ell_{\rm p}=\beta\kappa_{b,j}$). The interaction energy between the two strands (the Hydrogen bonding between two complementary bases) is modeled \textit{via} a Morse potential~\cite{PB} of width $\lambda$ and depth $A$:
\be
\mathcal{H}_{\rm int} = \sum_{j=1}^{N}A\left(e^{-2\frac{\rho_j-\rho_0}{\lambda}}-2 e^{-\frac{\rho_j-\rho_0}{\lambda}}\right)
\ee
where $\rho_j=|\br^{(1)}_j-\br^{(2)}_j|$ is the distance between complementary bases at position $j$ along the chain and $\rho_{0}=1$~nm is the equilibrium distance. The stacking interaction is modeled by a bending modulus $\kappa_{b}$ which depends on $\rho$, interpolating from $\kappa_{\rm ds}/2=75\,k_{\rm B}T$ for dsDNA state to $\kappa_{\rm ss}=3\,k_{\rm B}T$ for single stranded one, according to~\cite{sung}
\be
\kappa_{b,j}= \frac{\kappa_{ds}}2-\left(\frac{\kappa_{ds}}2-\kappa_{ss}\right)\,f(\rho_{j-1})f(\rho_j)f(\rho_{j+1})
\ee
where $f(\rho_j) = [1+\mathrm{erf}(\frac{\rho_j-\rho_b}{\lambda'})]/2$, $\lambda'$ is the width of the transition and $\rho_b=1.5\rho_0$. The variable bending modulus depends on three consecutive base-pair distances, which provides cooperativity. We chose $\lambda = 0.2$~nm, $\lambda' = 0.15$~nm, $\rho_0 = 1$~nm, and $\beta A=8$.\footnote{This value is chosen such that an initial dsDNA remains always closed in the longest simulation run.} The threshold value for $\rho$, discriminating between open and closed states, is fixed at 1.13~nm (a slightly different value does not change the results). 

The evolution of $\br_i(t)$ is governed by the overdamped Langevin equation
\be
\zeta \frac{\md\br_i}{\md t} = -\nabla_{\br_{i}}\mathcal{H}(\{\br_j\})+\bxi_{i}(t) 
\ee
where $\zeta=3\pi\eta a$ is the friction coefficient for each bead of diameter $a$ ($\eta=10^{-3}$~Pa.s is the water viscosity), $\bxi_{i}(t)$ is the random force (with zero mean), which mimics the action of the thermal heat bath and obeys the fluctuation-dissipation relation $\langle \bxi_i(t)\cdot\bxi_j(t') \rangle = 6k_{\rm B}T\zeta\,\delta_{ij}\,\delta(t-t')$. The adimensional time step, $\delta \tau=\delta t k_{\rm B}T/(a^2\zeta)$, was fixed, for sufficient accuracy, at $5\times10^{-4}$ ($\delta t=0.045$~ps). The initial bubble is created by turning off $\mathcal{H}_{\rm int}$ and then equilibrated for $2\,\mu$s. Output values are then calculated every $10^3$ steps once $\mathcal{H}_{\rm int}$ is turned on, and total simulation times range between $10^7$ to $10^8$ steps (0.4 to $4\,\mu$s). Samples are made of 200 trajectories and error bars are standard deviations.

\subsection{Kinetic Monte Carlo simulations} 

Out-of-equilibrium dynamics of the mean DNA chain (the center of mass of the two strands) has also been explored numerically by KMC simulations. We implemented the coupled model defined in Refs.~\cite{PRL,JPCM} where the mean chain is composed of $N$ identical beads representing the base-pairs. Simulation details are given in Ref.~\cite{PhysBiol} (each bead now represents one base-pair and has the mobility of a pair of beads in BD simulations). At each Monte Carlo Sweep of physical duration $\delta t=0.019$~ps, a bead is chosen at random and a random move is attempted for this bead. In addition, at each Monte Carlo step, we also attempt to flip the sign of one Ising spin variable $\sigma_i$, according to a standard Metropolis procedure. However, it might be that, in a real DNA, the frequency of change of internal degrees of freedom is different from this arbitrarily chosen one. To rule out this possibility, we simulated various systems where $10^{-3}$ to 100 spin-flips are attempted per $\delta t$. The average closure times then changed by at most $\pm 20$\% as compared to 1 spin-flip, thus proving that this is not a critical issue. The good matching between the KMC and BD results below supports this observation.

\section{Closure dynamics}
\begin{figure}[t]
\hspace{1.4cm}\rotatebox{90}{\includegraphics[width=1.17cm]{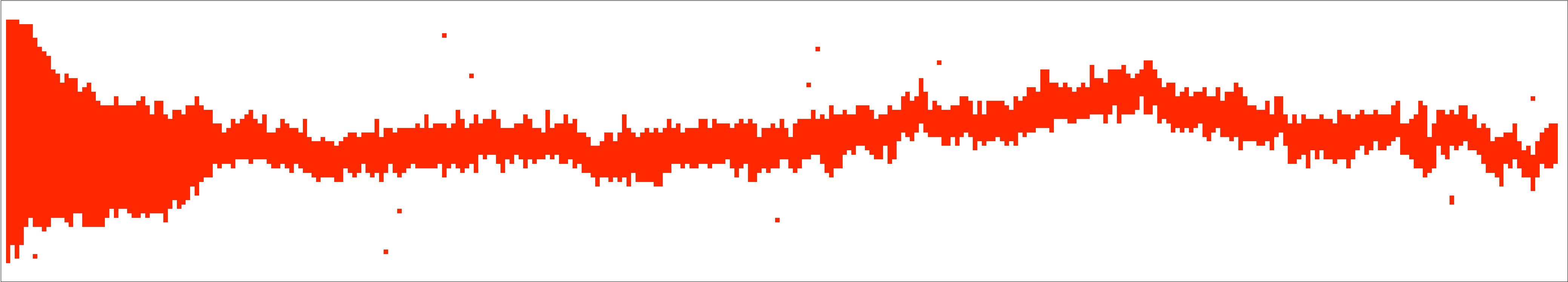}}
\includegraphics[width=8.5cm]{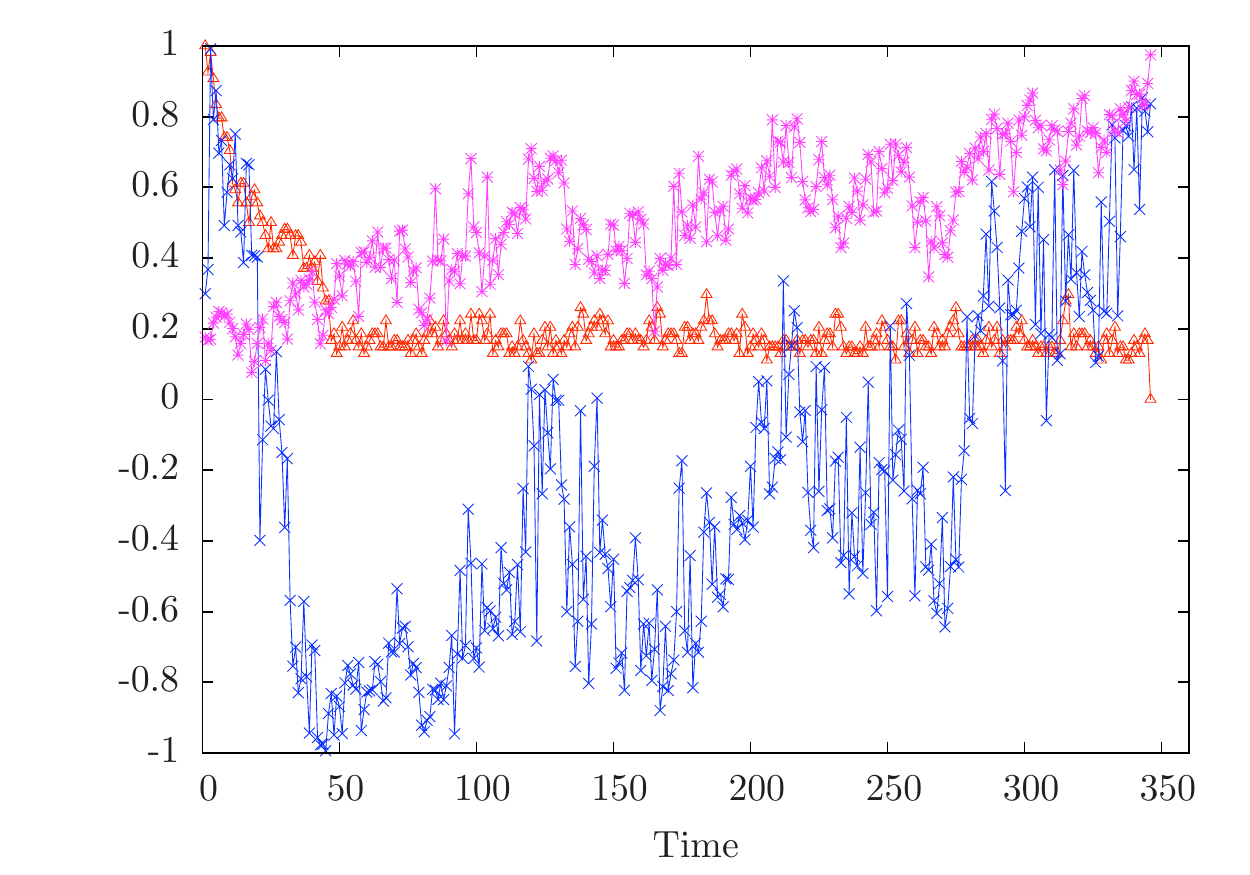}
\caption{Top: melting map (position of open base-pairs in DNA as a function of time). Bottom: time evolution of the bubble length, $L(t)/L_0$, in red, the bubble end-to-end distance, $R(t)/(aL(t))$, in purple, and the tangent-tangent product $C(t)$ in blue ($N=60$, time is in units of $10^4$ BD steps $=0.4$~ns).}
\label{typicalrun}
\end{figure}
The evolution of the bubble size (normalized by the initial bubble size $L_0$), $L(t)/L_0$, is plotted as a function of time in Figure~\ref{typicalrun} for a typical simulation ($N=60$). Two regimes appear clearly: at short times, $L(t)$ decreases rapidly which corresponds to a \textit{zipping} dynamics, until it reaches a \textit{metastable state} characterized by a small bubble of constant size $\bar L\simeq 10$, the center of which diffuses slowly (cf. Fig.~\ref{typicalrun} top). The simulation is stopped when the bubble closes, which defines the Mean First Passage Time (MFPT) for closure. The dimensionless bubble end-to-end distance $R(t)/aL(t)$ and the tangent-tangent product $C(t)\equiv\bt_{\rm init}\cdot\bt_{\rm end}$, where $\bt_{\rm init}$ (resp. $\bt_{\rm end}$) is the mean value of the two strand tangent vectors at the beginning (resp. end) of the bubble (see Fig.~\ref{snapshots}), are also plotted. First, $R(t)<aL(t)$ during the entire simulation run, except at closure when $R(t)\simeq aL(t)$ which favors the formation of dsDNA which is stiff at this length scale. Likewise, $C(t)$ starts here from a positive value and then decreases to a negative value in the metastable state close to $-1$, which corresponds to anti-correlated tangent vectors. It then undergoes large fluctuations, and the bubble closure corresponds to a value of $C(\tau_{\rm closure})\approx1$, i.e. when the two stiff arms are aligned. This type of behaviour is observed whatever the initial condition $C(0)$. The zipping is faster than the diffusion time of the small arms so that the distance between both DNA extremities remains almost constant while the bubble ``pushes'' in the direction parallel to the arms such that $C(\tau_{\rm zip})\simeq-1$. The DNA adopts an ``hairpin'' configuration as observed in Fig.~\ref{snapshots}b. The correlation between $C(t)$ and $L(t)$ is a clear indication that the spatial configuration of DNA plays a central role in bubble closure. 
\begin{figure}[t]
\includegraphics[width=8.5cm]{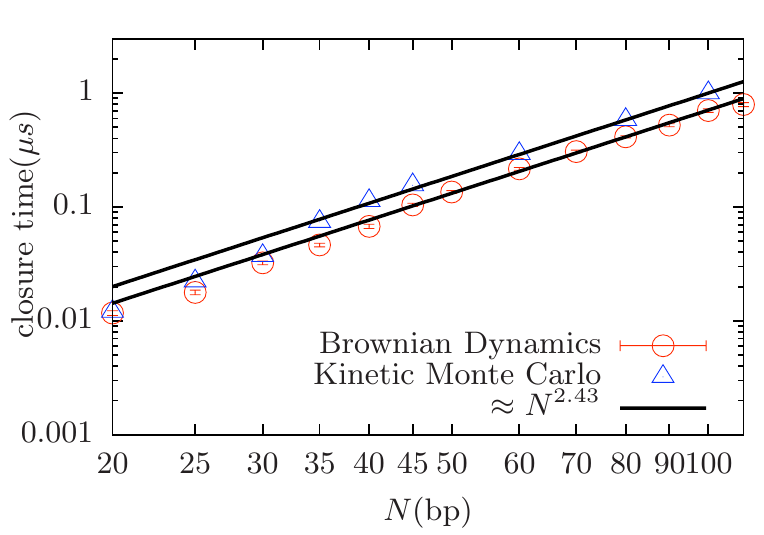}
\caption{Log-log plot of the total closure time, $\tau_{\rm closure}$, \textit{vs} total DNA length $N$. Dots corresponds to BD simulations results and triangles to KMC ones. The solid line is a fit between $N=50$ and 100 yielding $\tau_{\rm closure}\sim N^{2.43}$.}
\label{closure}
\end{figure}

Closure times simulated using both BD and KMC are shown in Figure~\ref{closure}. For one, the two types of simulations yield almost the same closure times and variation with $N$, showing that this quantity does not depend on the specific model. Since in KMC simulations the two strands are not simulated explicitly, this indicates that closure dynamics is dominated by the whole chain dynamics, which is the slowest process. For another, the fit of numerical data yields $\tau_{\rm closure}\sim N^{\alpha}$ with an apparent exponent $\alpha=2.4\pm0.1$ for the largest values of $N$ and almost $\alpha=3$ for small $N$ values. 
In any case an exponent $\alpha>2$ is found which is larger than $\alpha=1$ found in breathing dynamics using the Poland Scheraga model~\cite{metzlerJPA,metzlerPRL}, $\alpha=0.52$ in thermal renaturation~\cite{sikorav}, or $\alpha=1.37$ in anomalous zipping dynamics~\cite{carlon}. We argue below that this apparent exponent is the signature of a complex dynamics governed by the rotational friction of the (almost) rigid arms in the metastable state. We now discuss in details the two regimes.
\begin{figure}[t]
\includegraphics[width=8.5cm]{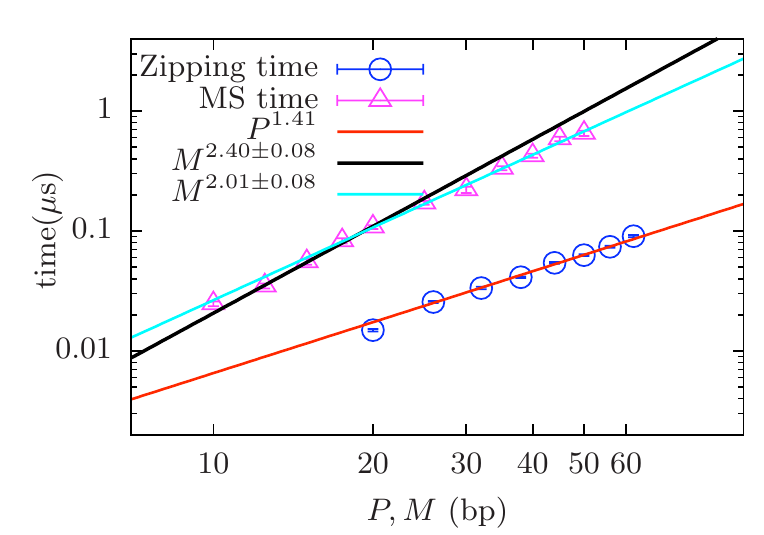}
\caption{Log-log plot of the zipping time (circles) and the metastable (MS) residence time (triangles) \textit{vs} the zipped ssDNA length $P=3(L_0-\bar L)/5$ and the arm's length $M$ respectively. Fits lead to $\tau_{\rm zip}\simeq P^{1.4}$ and $\tau_{\rm met}\sim M^2$ to $M^{2.4}$.}
\label{zipping}
\end{figure}
\subsection{Fast zipping process} In Figure~\ref{zipping} it is shown that the fast zipping process is characterized by an anomalous exponent for $L(t)$ and  the zipping time
\be
L_0-L(t)\sim t^{1/1.4},\qquad\tau_{\rm zip}\sim P^{1.4}
\label{zipping_exp}
\ee
required to zip $P\equiv\frac35(L_0-\bar L)$~bp. We also simulated DNA renaturation with an open end (data not shown) and found a zipping regime alone, with the same exponent.  A similar ``anomalous'' exponent ($\alpha\simeq 1.37$) has been found in a numerical study of the zipping dynamics of two polymers connected at one end~\cite{carlon}. This anomalous dynamics is interpreted, by analogy with polymer translocation~\cite{carlon,metzler}, as an out-of-equilibrium propagation of the tensile force located at bubble ends, $f\simeq A/a$, along the bubble backbone (the analog of the force applied on the monomer located at the tiny pore in polymer translocation)~\cite{sakaue}. The dynamics of $L(t)$ is governed by the equilibrium between friction, driving and bending forces along the chain 
\be
\eta \, b(t) \frac{\md(a L)}{\md t}=-f-\frac{\partial E_{\rm bend}}{\partial (aL)}
\label{zipping_eq}
\ee
where $b(t)$ is the bubble segment size involved in the friction, and $E_{\rm bend}$ is the bending energy stored in the bubble. Since the initial bubble is pre-equilibrated, $E_{\rm bend}$ can be neglected and the initial zipping is governed by the two first terms of \eq{zipping_eq}. 
For a fully flexible bubble, it has been proposed~\cite{kardar,grosberg,sakaue} that $b(t)\simeq aL(t)^\nu$ (where $\nu$ is the Flory exponent) which yields $\tau_{\rm zip}\simeq P^{1+\nu}/f$, an exponent value slightly larger than 1.4. Note that (i)~as the tensile force straightens the bubble, the bending force increases thus defining a  moderate forcing regime. In this regime ($a/R_0<\beta af<1$), Sakaue found, for the translocation case~\cite{sakaue}, an exponent value of 1.43  which agrees well with \eq{zipping_exp}. (ii)~Moreover, at the end of zipping, the ssDNA is not fully flexible since $\bar L \approx 2\ell_{\rm ss}$, which slightly increases the friction and thus the anomalous exponent.
Finally, the bubble geometry imposes that the zipping process stop when the bubble is highly bent. From \eq{zipping_eq} the metastable state is reached when $E_{\rm bend}(\bar L)\simeq f a \bar L$. For a (circular) bent bubble, $E_{\rm bend}(L)= 2\pi^2 2\kappa_{ss}/L$ which leads to $\bar L\simeq \sqrt{4\pi^2\kappa_{\rm ss}/A}$, i.e. a few base-pairs. In other words, when $L(t)=\bar L$, the bending energy cost for closing one more bp becomes larger than the base-pairing gain.

\subsection{Diffusion limited closure of the metastable state} 
\begin{figure}[t]
\includegraphics[width=8.5cm]{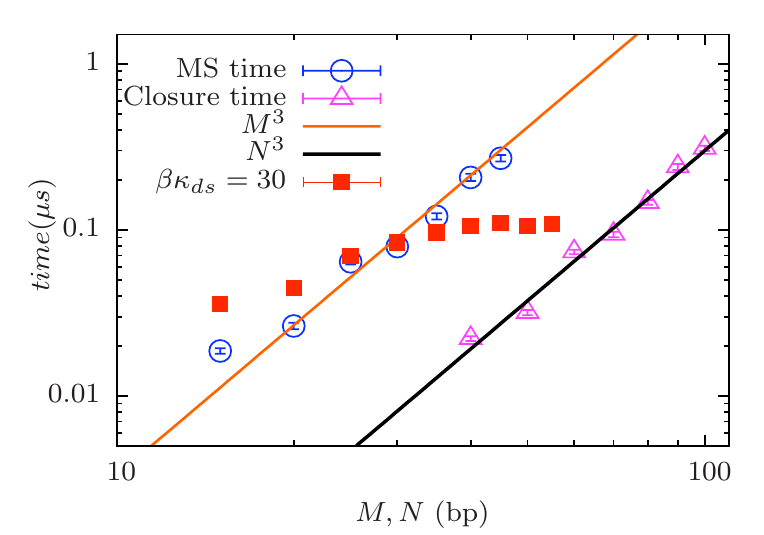}
\caption{Log-log plot of (i)~the metastable residence time and the closure time \textit{vs} the arm's length $M$ (circles) and chain length $N$ (triangles) respectively for $\beta\kappa_{\rm ds}=400$.  Fits lead to $\tau_{\rm met}\simeq M^{3}$ and $\tau_{\rm closure}\sim N^3$. (ii)~The metastable residence time \textit{vs} $M$ for $\beta\kappa_{\rm ds}=30$ (squares) showing a saturation for $M>2\beta\kappa_{\rm ds}$.}
\label{stifferdsDNA}
\end{figure}

The metastable residence time $\tau_{\rm met}$ is plotted in Figure~\ref{zipping} as a function of the arm size, $M$. Clearly a scaling law $\tau_{\rm met} \simeq M^2$ appears for the longest arms. For shorter (and thus stiffer arms), the exponent is larger, around 2.4. 
This law might be  surprising since $\beta\kappa_{\rm ds}/M\simeq 3-10$ and one would expect the arms to be stiff. A scaling estimate for two stiff arms connected by a flexible joint yields a time inversely proportional to the rotational diffusion coefficient of a rodlike polymer of length $M$ (neglecting prefactors and logarithmic terms coming from hydrodynamics)~\cite{doi}
\be
\tau_{\rm R}\simeq D_{\rm R}^{-1}\simeq \frac{\eta (aM)^3}{k_{\rm B}T}
\label{stiff_rod}
\ee
To check this law numerically and since we are limited in $M$ due to computational cost, we performed BD simulations with a larger dsDNA persistent length, $\beta\kappa_{\rm ds}=400$. Figure~\ref{stifferdsDNA} shows that both the metastable residence time and the closure time now scale like $\tau_{\rm met}\simeq M^{3}$ and $\tau_{\rm closure}\sim N^3$ respectively. This thus demonstrates that the closure process is limited by the rotational diffusion of the arms.
The reason why the exponent deviates from 3 to 2 when we consider a real DNA with $\beta\kappa_{\rm ds}=150 $ is twofold. First, the arms being semi-flexible, bending modes enter into play which decrease slightly the friction and accelerate the dynamics. Second, the rotational diffusion law \eq{stiff_rod} is only valid at long time scales and the closure may occur before the diffusive regime is reached.

\begin{figure}[t]
\includegraphics[width=8.5cm]{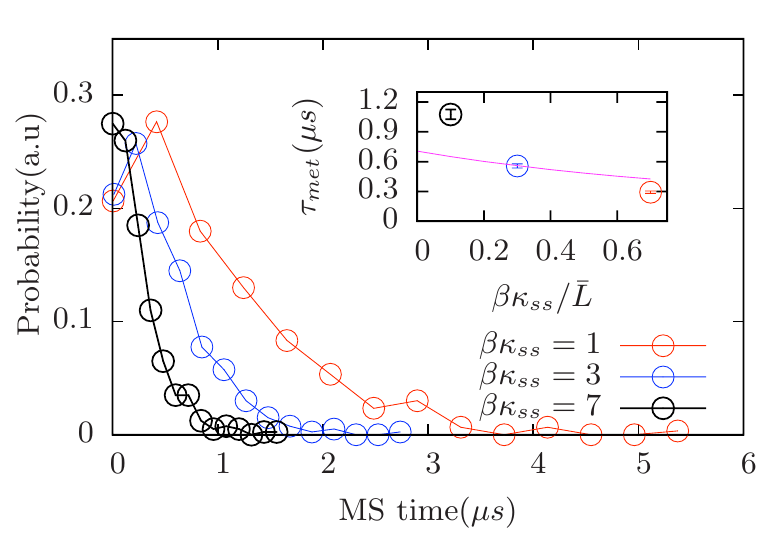}
\caption{Closure time probability for $\beta\kappa_{\rm ss}=1,3$  and 7 (300, 600 and 400 samples respectively) and $N=100$. The average value, $\tau_{\rm met}$, is plotted in the inset and clearly decreases with increasing $\kappa_{\rm ss}$. The solid line corresponds to \eq{MFPTtorque}.}
\label{distribution}
\end{figure}

To clarify this behaviour, we model the system as follows. The dynamics of closure is characterized by two coordinates, namely the end-to-end distance of the two single strands forming the bubble of length $\bar L$, $R(t)$, and the angle between the tangent vectors located at bubble ends, $\theta(t)=\arccos C(t)$ (see Fig.~\ref{snapshots}). The metastable residence time, $\tau_{\rm met}(\br)$ [where $\br=(R,\theta)$], is defined as the MFPT needed to go from this metastable state to the state where $R\simeq 2a L$ and $\theta\simeq \theta_c\ll1$. It is solution of a backward Smoluchowski diffusion equation in a potential $U(\br)$~\cite{hanggi}
\be
-\beta \nabla U \cdot\nabla [D\tau_{\rm met}(\br)]+ \Delta [D\tau_{\rm met}(\br)]=-1
\label{backSmol}
\ee
and scales like $\tau_{\rm met} \sim D_{\rm R}^{-1}$ and $\tau_{\rm met} \sim R^2/D_{\rm T}$ for rotational and translational diffusion, respectively. Solving \eq{backSmol} is out of reach, and we first concentrate only on rotational diffusion. The potential $U(\theta)=\kappa_{\rm ss} (1-\cos\theta)/\bar L$, which favors the $\theta=0$ state, is the effective bending free energy of the bubble. We find
\bea
\tau_{\rm met}(\theta_c|\theta) = \frac{\bar L}{2\ell_{\rm ss} D_{\rm R}}\left[t\left(\cos\theta,\frac{\ell_{\rm ss}}{\bar L}\right)-t\left(\cos\theta_c,\frac{\ell_{\rm ss}}{\bar L}\right)\right]\nonumber \\
t(x,y) = \ln\left(\frac{1-x}{1+x}\right) +\mathrm{Ei}[-y(1+x)]-e^{-2y}\mathrm{Ei}[y(1-x)]\label{MFPTtorque}
\eea
where $\mathrm{Ei}[z]=-\int_z^\infty \frac{e^{-t}}{t}\md t$. In the limit of an infinitely flexible bubble, $\bar L/\ell_{\rm ss}\to\infty$, \eq{MFPTtorque} simplifies into
\be
\tau_{\rm met}^\infty(\theta_c|\theta_0)=\frac2{D_{\rm R}}\ln\left[\frac{\sin(\theta_0/2)}{\sin(\theta_c/2)}\right]
\label{MFPT}
\ee
Simulation data show that, in the metastable state, the probability distribution of $\theta_0$ has a maximum around 2.5~rad. By choosing a small value such as $\theta_c=\pi/10$ as observed and $\bar L\simeq 2\ell_{\rm ss}$, one finds $\tau_{\rm met}\simeq \tau_{\rm met}^\infty/2$. \eq{MFPTtorque} reproduces qualitatively the decrease of $\tau_{\rm met}$ when $\kappa_{\rm ss}$ increases, as observed in Fig.~\ref{distribution} for $\beta\kappa_{\rm ss}=1$,~3 and 7. However, the inset shows that the argument leading to \eq{MFPTtorque} does not reproduces quantitatively this variation. 

The diffusion of the bubble should also be taken into account. We discuss two limiting cases: (i)~a stiff bent bubble which corresponds to a maximal correlation between bubble and arm rotational diffusion. For a circular bubble, one has geometrically $R=2aL\sin(\theta/2)/\theta$ and the metastable residence time is still given by \eq{MFPTtorque}. (ii)~The opposite, fully flexible bubble ($\ell_{\rm ss}\ll \bar L$) where the two diffusion processes are uncorrelated. The metastable residence time is roughly the longest time of (i)~$\tau_{\rm met}^\infty$ defined in \eq{MFPT} and (ii)~the MFPT necessary to go from the initial size $R(\tau_{\rm zip})$ to an almost extended configuration $R\simeq a \bar L$, which for an entropic spring with $\beta U(R)=R^2/(Na^2)$, is the Kramers-like result~\cite{hanggi}:
\be
\tau\simeq \eta \beta a^3 \bar L(4\bar L+M) \int_1^{\bar L^{1/2}}\md x\int_0^x \md y\left(\frac{y}{x}\right)^2\,e^{x^2-y^2}
\ee
where both bubble and arm translational frictions come into play. As soon as $\bar L\simeq 10<M$ this MFPT is smaller than $\tau^\infty_{\rm met}$. In the real case of a semi-flexible bubble ($\bar L\simeq\ell_{\rm ss}$), the two diffusive processes are entangled, which yields larger closure times in $N^{2.4}$, as observed in the simulations.

Finally, although simulations were done for stiff arms with $M\leq\beta\kappa_{\rm ds}$, we have checked by decreasing $\beta\kappa_{\rm ds}$ from 75 to 30, that for long semi-flexible arms, $\tau_{\rm met}(M)$ saturates for $M\geq 2\beta\kappa_{\rm ds}$, as shown in Fig.~\ref{stifferdsDNA}. Indeed, to close the bubble it is sufficient that the two segments of length $\simeq\ell_{\rm ds}$ adjacent to the bubble align since the rest of the dsDNA chain is decorrelated. This yields an upper limit for the diffusion time of $\tau_{\rm met}^{\max}\simeq \eta \beta (2a\ell_{\rm ds})^3 \simeq 20\;\mu$s for $\ell_{\rm ds}=150$~bp.

\section{Discussion}

Using Brownian dynamics simulations of a coarse-grained model of DNA, we show that the closure of a pre-equilibrated denaturation bubble of size larger than 10~bp is dominated by the rotational diffusion of the stiff dsDNA arms. The first stage consists in a fast zipping, the driving force of which is the base pairing energy on the order of $8\,k_{\rm B}T$~\cite{PRL}, leading to an average growth rate of $10^7$~bp/s (even though the time dependence of the bubble size is not linear). This fast zipping regime stops when the bending energy stored in the small bubble is measured to be roughly $20\,k_{\rm B}T$. The closure is then diffusion-controlled with a small driving force due to the bending torque applied by the bubble. We numerically find closure times $\tau_{\rm closure}\sim N^{2.4}$ for DNA lengths smaller than the dsDNA persistence length of 150~bp. For larger DNA lengths, we show that the closure time saturates at $20\;\mu$s.

Our key conclusion is that DNA closure dynamics is governed by chain bending, which is a 3 dimensional effect that cannot be accounted for by 1D models (such as Poland-Scheraga~\cite{PS,metzlerJPA,metzlerPRL} or Peyrard-Bishop~\cite{PB} models). Aside from small time scales, much smaller than the Rouse time of the DNA chain (where DNA breathing dynamics is governed by the opening/closure of base pairs), DNA bubble dynamics involves several intertwined ingredients such as base-pair closure, DNA chain diffusion, and elastic forces.

Taking into account hydrodynamic interactions would slightly decrease the apparent exponent of the scaling laws. For instance, the rotational diffusion time of a rigid rod, given in \eq{stiff_rod}, would be divided by a term in $\ln M$~\cite{doi}. Hence as soon as small and thus stiff DNAs are considered, the correction to the free draining case remains small.
We did not consider the helical structure of DNA in this work. In recent numerical work~\cite{baiesi}, Baiesi \textit{et al.} show that the unwinding dynamics of a helical polymer made of two interwound strands (without any attractive interaction) leads to a unwinding time which scales as $N^{2.57}$. One might expect that the closure of a denaturation bubble would include both winding of the bubble plus rotational diffusion of the two arms in order to relax the superhelical stress~\cite{benham}. The precise role of the axial rotational dynamics of the two interwound strands is currently under study. A lower bound for the closure time can be estimated by considering the winding of an initial bubble of  length $L_0$. The stochastic evolution of a winding angle $\phi(t)=2\pi L(t)/p$ (where the DNA pitch is $p=10.5$~bp) is given by $\zeta_{\rm rot}(t) \md \phi/\md t=n(t)$ where $n(t)$ is a random torque along the DNA axis with $\langle n(t) n(t')\rangle=2 k_{\rm B}T \zeta_{\rm rot}(t)\delta(t-t')$. Since $\zeta_{\rm rot}\sim L(t)^{z\nu}$ where $z\nu=2,\frac32,\frac95$ (Rouse, Zimm models in theta and good solvent respectively~\cite{doi}), one finds after integration 
\be
\tau_{\rm closure}\sim L_0^{2+z\nu}\sim N^{2+z\nu}
\ee
where $3.5\leq2+z\nu\leq4$. Hence, DNA winding might be the missing mechanism explaining the quantitative gap between our estimate of the closure time of $0.04\;\mu$s for a 30~bp long DNA (see Fig.~\ref{closure}) and values of $1-100\;\mu$s measured experimentally~\cite{altan,warmlander}. 
\\

The authors acknowledge Cl\'ement Chatelain who performed some simulations during his Master thesis.

\end{document}